\documentclass[aps,prl,twocolumn,superscriptaddress]{revtex4-1}
\usepackage{mathrsfs}
\usepackage{physics}
\usepackage{hyperref}
\usepackage{bm}
\usepackage{url} 
\usepackage{graphicx}
\usepackage{subfigure}
\usepackage{xcolor}
\usepackage{bm}
\usepackage{bbm}
\usepackage{times}
\usepackage{stix}
\usepackage{soul}
\usepackage[normalem]{ulem}

\newcommand{\ba}{\begin{equation}\begin{aligned}}
\newcommand{\ea}{\end{aligned}\end{equation}}
\newcommand{\sct}[1]{\cite{#1}}

\makeatletter

\begin{document}
\title{Intrinsic interface adsorption drives selectivity in atomically smooth nanofluidic channels}
\author{Phillip Helms}
\affiliation{Department of Chemistry, University of California, Berkeley, California 94720, USA}
\affiliation{Chemical Science Division, Lawrence Berkeley National Laboratory, Berkeley, California 94720, USA}
\author{Anthony R. Poggioli}
\affiliation{Department of Chemistry, University of California, Berkeley, California 94720, USA}
\affiliation{Kavli Energy NanoScience Institute, Berkeley, California 94720, USA}
\author{David T. Limmer}
\email{dlimmer@berkeley.edu}
\affiliation{Department of Chemistry, University of California, Berkeley, California 94720, USA}
\affiliation{Chemical Science Division, Lawrence Berkeley National Laboratory, Berkeley, California 94720, USA}
\affiliation{Materials Science Division, Lawrence Berkeley National Laboratory, Berkeley, California 94720, USA}
\affiliation{Kavli Energy NanoScience Institute, Berkeley, California 94720, USA}

\date{\today}

\begin{abstract}
Specific molecular interactions underlie unexpected and useful phenomena in nanofluidic systems, but require descriptions that go beyond traditional macroscopic hydrodynamics. 
In this letter, we demonstrate how equilibrium molecular dynamics simulations and linear response theory can be synthesized with hydrodynamics  to provide a comprehensive characterization of nanofluidic transport.
Specifically, we study the pressure driven flows of ionic solutions in nanochannels comprised of two-dimensional crystalline substrates made from graphite and hexagonal boron nitride. 
While simple hydrodynamic descriptions do not predict a streaming electrical current or salt selectivity in such simple systems, we observe that both arise due to the intrinsic molecular interactions that act to selectively adsorb ions to the interface in the absence of a net surface charge. 
Notably, this emergent selectivity indicates that these nanochannels can serve as desalination membranes. 
\end{abstract}
\maketitle

Recent advances in nanoscale fabrication techniques have enabled the synthesis of nanofluidic systems with novel functionalities, \sct{bocquet2010nanofluidics, bocquet2020nanofluidics, eijkel2005nanofluidics} with applications to biotechnology~\sct{segerink2014nanofluidics}, filtration~\sct{gao2017nanofluidics, zhang2021nanofluidics, kim2010direct}, and computation~\sct{robin2021modeling, hou2021bioinspired, sheng2017transporting}.
For example, nanofluidics-based membranes have leveraged atomic level details like those of evolved biological membranes
~\sct{hub2008mechanism, murata2000structural, chen2022engineering, 
zhang2020engineering, park2014carbon, 
schoch2008transport, daiguji2010ion, siria2017new} 
to circumvent traditional trade-offs between permeability and selectivity that plague membrane technology
~\sct{park2017maximizing,robeson2008upper,robeson1991correlation,beyond_the_tradeoff}. 
While continuum-level hydrodynamic descriptions can remain accurate at scales of a few nanonmeters, enabling some general design principles to be deduced~\sct{zhou2021wall, bocquet1994hydrodynamic, bocquet2007flow, chen2015determining}, 
the continued development of nanofluidic devices is limited by a lack of understanding of emergent interfacial effects which are resolutely molecular in origin. With large surface to volume ratios, the properties of fluids confined to nanometer scales are determined in large part by a delicate interplay of interactions between the bounding surfaces and the working fluid. To understand and design nanofluidic devices, an approach that combines macroscopic and molecular perspectives is necessary~\sct{limmer2021large}.

In this letter, we show how interfacial atomic structure affects the directed transport of an electrolyte solution in nanochannels made of atomically flat graphite (GR) and hexagonal boron nitride (BN) walls using molecular dynamics simulations unified with a contemporary perspective on hydrodynamics.
These simple nanofluidic systems have been studied extensively because of their intriguing transport properties,
such as anomalously high permeabilities in GR
~\sct{poggioli2021distinct, yang2018rapid, hummer2001water,
keerthi2021water, secchi2016massive, falk2010molecular, neek2018fast, 
tocci2014friction, tocci2020ab,poggioli2021distinct}, 
and the potential to augment their functionality with selectivity for desalination or blue energy applications
~\sct{boretti2018outlook, ang2020review, sun2016recent, li2020water, cohen2012water, o2014selective, liu2018water, montes2022ionic, mi2014graphene,joly2021osmotic}.
By computing the spatially-resolved volumetric, charge, and species transport coefficients from equilibrium correlations~\sct{mangaud2020sampling, agnihotri2014displacements, viscardy2007transport}  we elucidate  the importance of specific molecular interactions on nanofluidic device functionality.
While from a continuum perspective, driving the solution with a pressure gradient should result in salt filtration or electric current only when the confining walls have a net charge, we discover that the intrinsic interfacial adsorption of ions can lead to streaming electrical currents and a novel, emergent desalination mechanism.  


\begin{figure}[t]
\includegraphics[width=8.5cm]{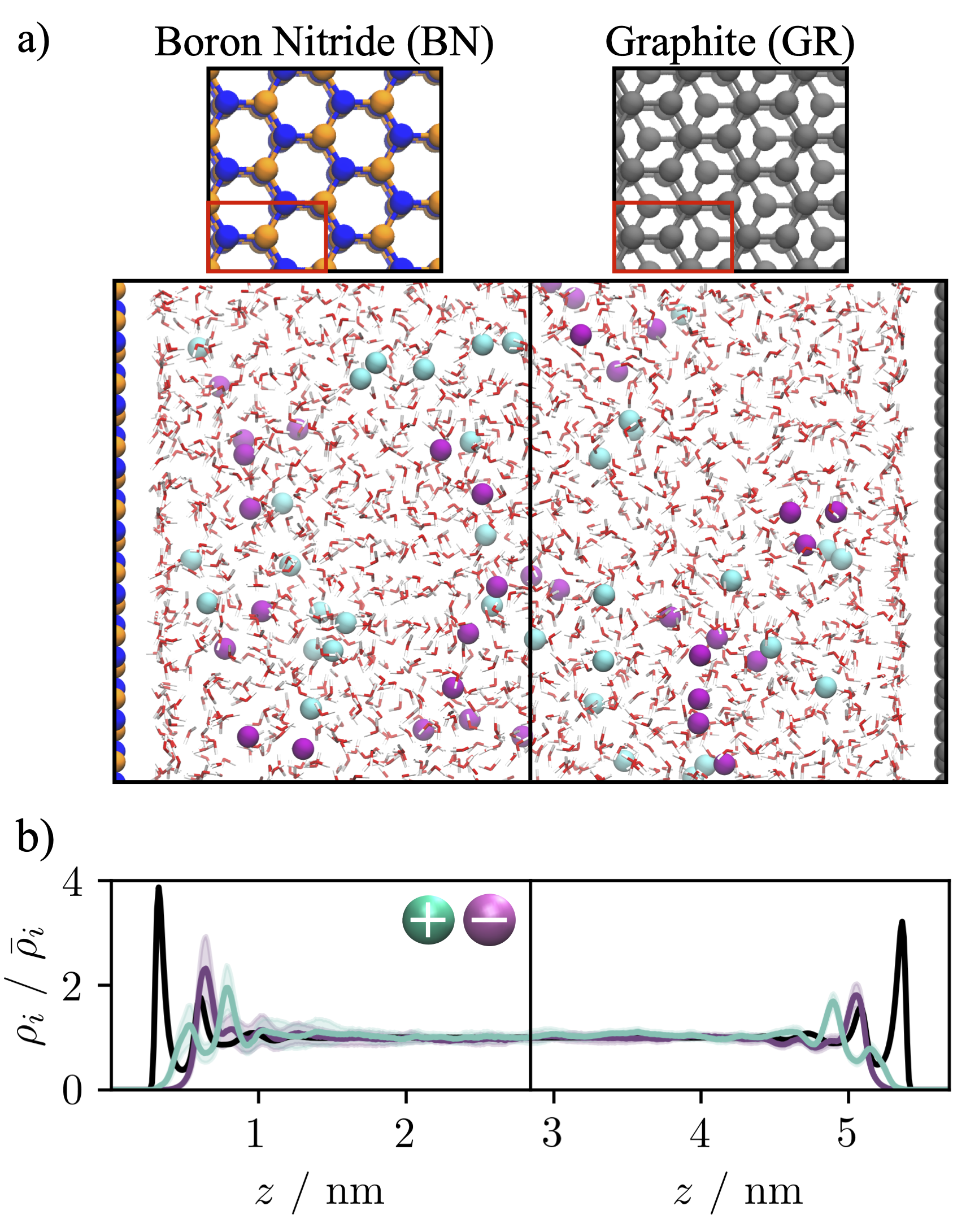}
\caption{
Description of the systems considered and resulting equilibrium 
density distributions. 
(a) A snapshot of the nanochannels considered with the left (right) side 
corresponding to the boron nitride (graphite) nanochannel. The top images
show the wall structure, with each wall composed of three layers and 
the periodic unit cell outlined in red. 
(b) The molecular species density distributions for potassium (green), chloride
(purple), and water (black) as a function of
position, normalized by bulk densities.}
\label{fig:system}
\end{figure}

We focus on the two systems illustrated in Fig.~\ref{fig:system}(a), consisting of 
an aqueous solution of potassium chloride confined in nanochannels with fixed walls of either 
BN or GR.
Because of the experimental similarity between the structure of BN and GR lattices, 
we spaced atoms and lattice layers identically, with interatomic and interlayer spacings of 1.42 \r{A}
and 3.38 \r{A}~\sct{solozhenko1995isothermal, ooi2006density}.
Each wall has three layers, 
using AA' and AB stacking for BN and GR, respectively, to match their equilibrium structures, 
with lattice unit cells repeated 8 and 13 times in the 
$x$ and $y$ directions for a cross-sectional surface area of nearly $ 9 \ \mathrm{nm}^2$.
The walls were separated such that the spacing between the center of mass of the 
innermost wall layers was $H\approx 5.7 \ \mathrm{nm}$, with the channel width adjusted to ensure
a bulk water density of $\bar{\rho}_\mathrm{w} \approx 1 \ \mathrm{g}/\mathrm{cm}^3$. 
The channels were filled with $N_\mathrm{w}=1920$ TIP4P/2005 water molecules
with rigid geometries imposed using the SHAKE algorithm~\sct{abascal2005general,ryckaert1977numerical},
$N_{\mathrm{K}^+}=40$ potassium ions and $N_{\mathrm{Cl}^-}=40$ chloride ions, resulting in a nearly 1 M electrolyte solution.

We evolved this system according to underdamped Langevin dynamics,
\ba\label{eq:langevin}
m_i\dot{\mathbf{v}}_i=
-\zeta_i\mathbf{v}_i
+\mathbf{F}_i\left(\mathbf{r}^N\right)
+\mathbf{R}_i
\ea
where each particle $i$ has mass $m_i$, 
velocity $\mathbf{v}_i$, and experiences a friction $\zeta_i$, 
with forcing from interparticle interactions
$\mathbf{F}_i\left(\mathbf{r}^N\right)$,
and random noise $\mathbf{R}_i$.  
The random force is a Gaussian random variable with mean 
$\langle R_{i,\alpha}\rangle=0$ and variance 
$\langle R_{i,\alpha}(t)R_{i',\alpha'}(t')\rangle=
2k_\mathrm{B}T\zeta_i\delta_{i,i'}\delta_{\alpha,\alpha'}\delta(t-t')$ 
for each cartesian coordinate $\alpha$, where $k_\mathrm{B}T$ is Boltzmann's constant times temperature. 
Periodic boundary conditions were imposed in all three spatial dimensions, 
with a vacuum layer in the $z$ direction of $5 \ \mathrm{nm}$ to ensure no interaction between periodic
images of the channel. 
Intermolecular Lennard-Jones forces were chosen from literature-reported values to 
reproduce the solubility of ions in water and match the \emph{ab initio} equilibrium 
fluid structure in BN and GR nanochannels~\sct{yagasaki2020lennard,kayal2019water,kayal2019water},
with Lorentz-Berthelot mixing rules defining heteroatomic interactions.
Additionally, water molecules, charged ions, and the BN wall atoms interacted with Coulomb potentials, where boron and nitrogen atoms have charges of $\pm \ 1.05 \mathrm{e}$, with $\mathrm{e}$ being the elementary charge, using an Ewald summation as implemented in LAMMPS~\sct{plimpton1995fast}.
For all data presented here, we performed 5 independent simulations, each starting with an equilibration 
run for $5 \ \mathrm{ns}$ with $m_i/\xi_i=2 \ \mathrm{ps}$, followed by a production run 
for $10-20 \ \mathrm{ns}$ with $m_i/\xi_i=10 \ \mathrm{ns}$ at a temperature of $298 \ \mathrm{K}$. 
In all plots, lines represent averages and error bars represent the standard deviation for the 5 simulations. 
All scripts used to produce these results and the raw data are openly available~\sct{zenodo_data}.

Figure~\ref{fig:system}(b) shows the equilibrium particle number densities, $\rho_i(z)$, for all species, $i=\{\mathrm{w},\mathrm{K}^+,\mathrm{Cl}^-\}$, in the BN and GR channels, relative to their bulk values, $\bar{\rho}_i$.
We observe similar structures in both materials with interfacial layering of water that is consistent with previous simulations of neat water\sct{kayal2019water, tocci2014friction}.
The distribution of ions near such interfaces is known to be highly dependent on ion species, and the profiles shown are consistent with previous simulations~\sct{pykal2019ion, elliott2022electrochemical, dockal2019molecular}
A dense layer of pure water accumulates near the wall, with the molecules oriented 
such that they induce a small local negative charge.
The next layers are enriched in alternating concentrations of potassium and chloride ions, 
with depletion (accumulation) of water molecules accompanying potassium (chloride) enrichment. 
The two materials differ slightly, with a higher water density in the first layer of BN resulting in layering with higher amplitude in BN compared to GR, though in both systems the layering in the density decays to its bulk value for each species, $\bar{\rho}_i$, within 1.5 nm.

We consider fluxes induced by a pressure differential, $-\Delta P_x$, 
 imposed electrostatic potential drop $-\Delta \Phi_x$, or water chemical potential differential, $-\Delta \mu_x$,
with subscripts denoting application in the $x$ direction parallel to the walls,
and limit ourselves to small driving strengths. 
In this limit, linear response theory dictates that induced local fluxes are linearly dependent on driving forces,
\ba\label{eq:lr}
\begin{pmatrix}
q(z) \\ j(z) \\ d(z)
\end{pmatrix}
=
\begin{pmatrix}
\mathcal{M}_{qQ} & \mathcal{M}_{qJ} & \mathcal{M}_{qD} \\
\mathcal{M}_{jQ} & \mathcal{M}_{jJ} & \mathcal{M}_{jD} \\
\mathcal{M}_{dQ} & \mathcal{M}_{dJ} & \mathcal{M}_{dD} \\
\end{pmatrix}
\begin{pmatrix}
-\Delta P_x \\ -\Delta \Phi_x \\ -\Delta \mu_x
\end{pmatrix},
\ea
where $q(z)$ is the volumetric flow, $j(z)$ the charge flux, $d(z)$ the excess water flux, and the $\mathcal{M}_{aB}(z)$ are the spatially dependent mobilities. 
The excess water flux $d(z)$ represents the local water flux relative to what would be predicted from the bulk water density and the local total flux of water and ions, and it is considered here because it is particularly relevant for desalination. The diagonal elements of the mobility matrix link a given forcing directly to its conjugate flux -- e.g., $\mathcal{M}_{jJ}$ links the potential drop $-\Delta \Phi_x$ directly to the induced charge flux $j(z)$ -- while the off-diagonal elements are the so-called cross-terms linking, for example, an induced charge flux to an applied pressure differential. The total fluxes include the total volumetric flow $Q$, charge flux $J$, and excess water flux $D$. We index mobilities  by the local induced flux $a$ and total flux $B$ directly conjugate to a particular forcing. 

The local fluxes are defined microscopically as
\ba
q(z, t) &= \frac{H}{N} \sum_{i=1}^N v_{i, x}(t)\delta\left[z-z_i(t)\right] \\
j(z, t) &= \frac{1}{A_\mathrm{s}} \sum_{i=1}^N c_i v_{i, x}(t)\delta\left[z-z_i(t)\right] \\
d(z, t) &= \frac{1}{A_\mathrm{s}}  \sum_{i=1}^{N} v_{i, x}(t) \left (\delta_{i,\mathrm{w}}-f_\mathrm{w}^\mathrm{b} \right )\delta\left[z-z_i(t)\right]
\\
\ea
where particle $i$ has velocity $v_{i, x}(t)$ and position $z_i(t)$ at time $t$, a static charge of $c_i$, and $\delta_{i,\mathrm{w}}$ is a Kroniker delta that returns 1 if particle $i$ is a water molecule and is 0 otherwise. 
The bulk mole water fraction is defined as $f_\mathrm{w}^\mathrm{b}=N_\mathrm{w}^\mathrm{b} / N^\mathrm{b}$, where $N_\mathrm{w}^\mathrm{b}$ and 
$N^\mathrm{b}$ are respectively the average numbers of water molecules and all molecules in the bulk and $A_\mathrm{s}$ is the surface area associated with the fluid-wall interface. 
The spatial dependence can be integrated out by defining total fluxes, such as $Q=1/H\int_0^H dz \ q(z)$, with analogous definitions for $J$ and $D$.
Total channel conductivities can be evaluated as $\mathcal{L}_{AB}=1/H\int_0^H dz \ \mathcal{M}_{aB}(z)$, resulting in total flux 
linear response relations such as $Q=-\mathcal{L}_{QQ} \Delta P_x - \mathcal{L}_{QJ} \Delta \Psi_x - \mathcal{L}_{QD} \Delta \mu_x$.
While the integrated conductivities must obey Onsager reciprocal relations, $\mathcal{L}_{AB}=\mathcal{L}_{BA}$, mobilities are under no such constraint. It is possible for $\mathcal{M}_{aB}(z)\ne\mathcal{M}_{bA}(z)$.

Rather than attempting to calculate mobilities directly via nonequilibrium simulations, we use fluctuation-dissipation relations in order to obtain transport coefficients from equilibrium flux correlations \sct{mangaud2020sampling, agnihotri2014displacements, viscardy2007transport}. This allows us to avoid running separate nonequilibrium simulations for each term in the mobility matrix, and ensures the validity of linear response.
We adopt the Einstein-Helfand approach over the Green-Kubo method, 
as recent work has demonstrated its enhanced statistical efficiency~\sct{mangaud2020sampling}.
Mobilities are obtained as the long time slope of the correlation between
time-integrated local and global fluxes
\ba\label{eq:mobility}
\mathcal{M}_{aB} = \frac{V}{2 k_\mathrm{B} T} \lim_{t\to\infty} \frac{\kappa_{aB}(t)}{t}, 
\ea
with the correlation function
\ba
\kappa_{aB} = 
\ \int_0^t dt' \int_0^t dt'' \ \left\langle a(z, t')  \ B(t'')
\right\rangle,
\ea
volume $V=A_\mathrm{s}H$, and brackets representing an equilibrium average.
Similarly, conductivities can be obtained using correlations between global fluxes, 
$\mathcal{L}_{AB} = ( V / 2 k_\mathrm{B} T )\lim_{t\to\infty} K_{AB}(t) / t$
with $K_{AB}=
\int_0^t dt' \int_0^t dt'' \left\langle A(t')  B(t'')
\right\rangle
$.


\begin{figure}[t]
\includegraphics[width=8.5cm]{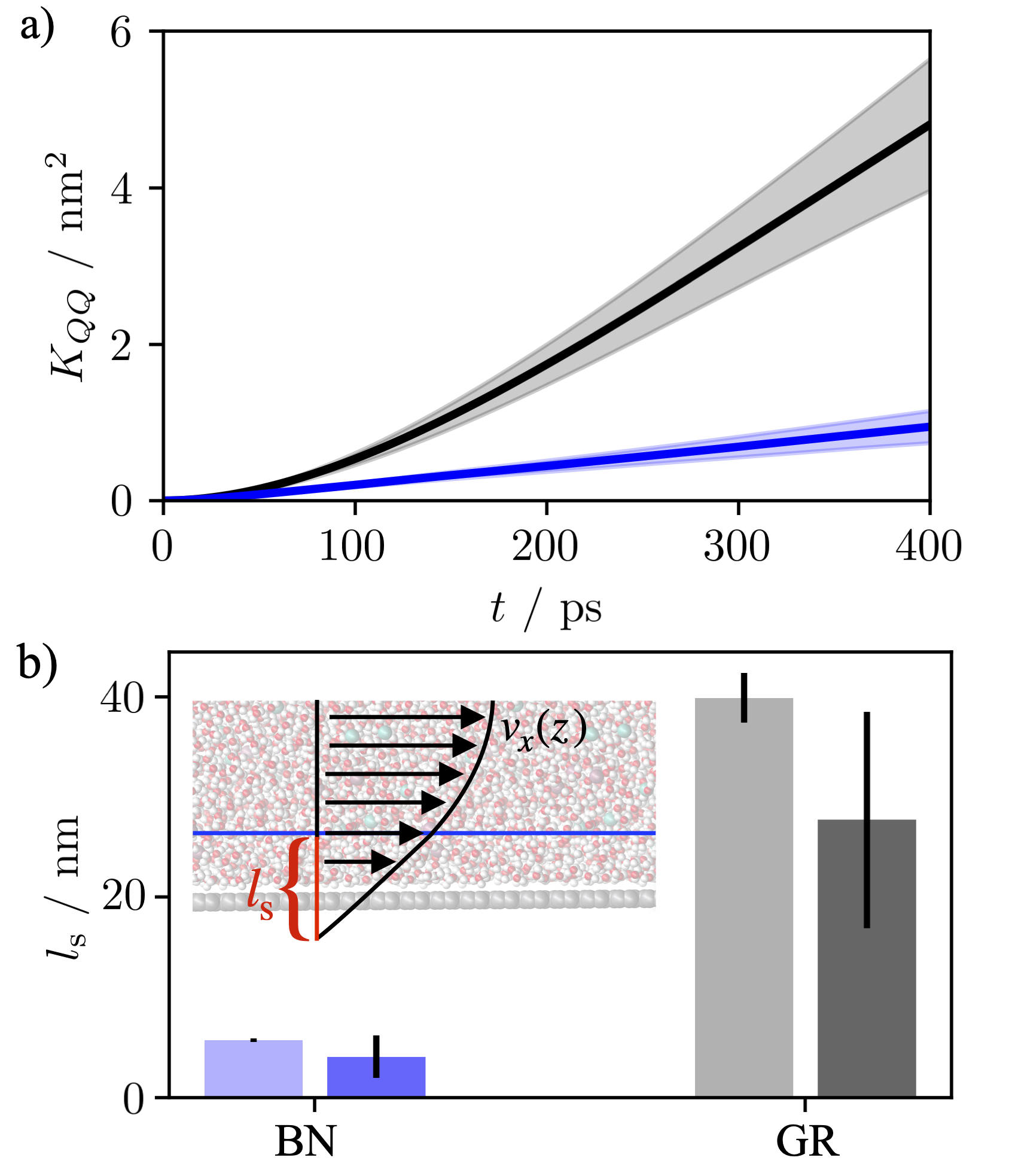}
\caption{
Comparison of the hydraulic conductivity and slip length for the 
GR (black) and BN (blue) nanochannels. 
(a) The time-integrated global flux 
correlation function $K_{QQ}$ versus time. 
(b)
Comparison of the slip lengths for both materials, computed from the hydraulic conductivity (dark), 
against previously reported results for neat water (light)
~\sct{poggioli2021distinct}. The inset illustrates the geometric interpretation of the slip length.
}
\label{fig:conductivity}
\end{figure}
Previous work has demonstrated that while equilibrium structures suggest only minor differences between 
water in BN and GR nanochannels, the dynamics of the confined fluid are strikingly different. This results in large differences in the friction between the fluid and walls, and significant differences in resultant channel permeabilities.
~\sct{poggioli2021distinct, tocci2014friction, thiemann2022water, Mouterde_et_al2019}.
In the presence of ions, the interfacial structure of water is altered and as a consequence the friction may change. 
In Fig.~\ref{fig:conductivity} (a), we show the integrated global flux correlation function $K_{QQ}$ as a 
function of time for both nanochannels. 
After approximately $200 \ \mathrm{ps}$, the correlation functions 
approach a linear dependence on time and their slopes give the hydraulic conductivities as
$\mathcal{L}_{QQ}^{\mathrm{[BN]}}=18.0\pm9.2 \ \mathrm{mol \ nm^5 \ kJ^{-1} \ ns^{-1}}$ and 
$\mathcal{L}_{QQ}^{\mathrm{[GR]}}=106\pm40 \ \mathrm{mol \ nm^5 \ kJ^{-1} \ ns^{-1}}$, which differ by nearly an order of magnitude.

While the hydraulic conductivities deduced above are independent of a specific hydrodynamic model, they can be connected to continuum theory through the slip length $l_\mathrm{s}$. In contrast to the no-slip condition typically applied in macroscopic contexts, which specifies that the fluid velocity exactly vanishes at the walls, the small confinement scales and enhanced importance of interfacial details in nanofluidic applications typically require application of the finite-slip condition. This condition specifies that the velocity at the wall is proportional to the shear strain at the wall, $v_x=l_\mathrm{s} (\partial v_x/\partial z)|_{z=0}$.
The slip length is interpreted geometrically as the distance beyond the interface where the 
extrapolated flow profile is zero, as illustrated in Fig.~\ref{fig:conductivity} (b).

To apply a hydrodynamic interpretation, we consider only the region where a 
hydrodynamic description is expected to be valid by defining the effective hydrodynamic interface 
as the the location of the second water density peak in Fig.~\ref{fig:system}(b)
~\sct{chen2015determining}. At this distance, microscopic density correlations have decayed and the fluid is well described as a continuous medium. 
The Poiseuille solution for the hydraulic mobility in the presence of a finite slip length is given by
\ba\label{eq:poiseulle}
\mathcal{M}_{qQ}(z)=\frac{H_\mathrm{hyd}^2}{2\eta}\left[\frac{l_\mathrm{s}}{H_\mathrm{hyd}}+\frac{z}{H_\mathrm{hyd}}-\frac{z^2}{H_\mathrm{hyd}^2}\right],
\ea
where $H_{\rm hyd}$ is the distance between hydrodynamic interfaces, and $\eta$ is the estimated viscosity of the solution. This expression may be integrated to determine the hydraulic conductivity
\begin{equation}
\label{eq:hyd_cond}
\mathcal{L}_{QQ} = \frac{H_{\rm hyd}^2}{12 \eta} \left( 1 + 6 \frac{l_{\rm s}}{H_{\rm hyd}} \right),
\end{equation}
which allows us to relate the measured values of $\mathcal{L}_{QQ}$ in GR and BN to the corresponding slip lengths provided $\eta$ is known.
Here, we use a viscosity of $\eta=1.0 \ \mathrm{mPa \ s}$, obtained by interpolating literature values for this electrolyte model~\sct{yagasaki2020lennard}.
Figure~\ref{fig:conductivity} (b) indicates the resulting slip lengths, 
$l_\mathrm{s}^{\mathrm{[BN]}}=4.0\pm2.5 \ \mathrm{nm}$ and $l_\mathrm{s}^{\mathrm{[GR]}}=27\pm10 \ \mathrm{nm}$, and compares them against previously reported results for neat water~\sct{poggioli2021distinct}. 
With the slip in GR nanochannels being approximately an order of magnitude larger than the slip in BN nanochannels, it is clear that the qualitative results do not change significantly with the addition of salt. 
The material-dependency of $l_\mathrm{s}$ has been observed in various contexts experimentally
~\sct{secchi2016massive, secchi2016scaling, holt2006fast,xie2018fast,majumder2005enhanced}
and is generally understood to arise from a decoupling of structure and dynamics, though the precise physical mechanism is debated
~\sct{faucher2019critical, poggioli2021distinct, thiemann2022water, bui2022classical, tocci2014friction,kavokine2022fluctuation}.
Quantitatively, our simulations also suggest a decrease in slip as salt is added, 
which is consistent with other observations for slip on hydrophobic surfaces, 
where increasing fluid-wall friction results as a consequence of enhanced equilibrium force fluctuations from the heterogeneous solution. 
~\sct{bakli2013effect, barrat1999influence, joly2004hydrodynamics}.

\begin{figure}
\includegraphics[width=8.5cm]{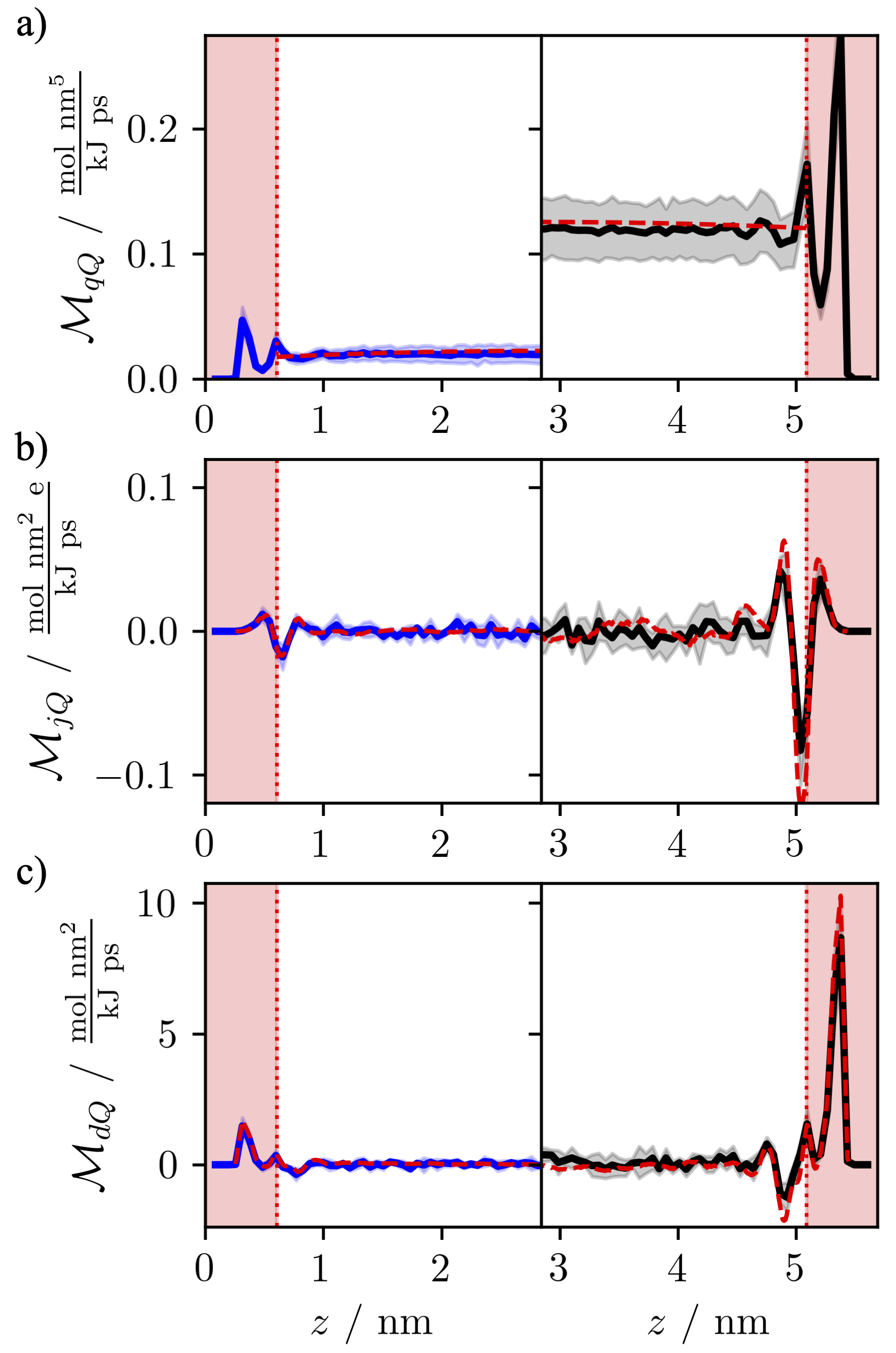}
\caption{Pressure driven hydraulic (a), streaming (b), and 
excess water (c) mobility profiles for BN (left, blue) and GR (right, black). 
The red shaded regions demarcate areas where hydrodynamics are invalidated. 
In (a), the red dashed curve corresponds to the hydrodynamic estimate
from the hydraulic conductivity. In (b) and (c), the red dashed curves 
are the mobility predictions from the product of the hydraulic mobility and 
appropriate density. 
}
\label{fig:mobilities}
\end{figure}
More detailed insight into the differences in transport characteristics between BN and GR nanochannels 
can be obtained by computing the spatially-dependent hydraulic mobility using Eq.~\ref{eq:mobility}. The results of this calculation for GR and BN are shown in Fig~\ref{fig:mobilities} (a).
We also show the hydrodynamic mobility profiles calculated from Eq.~\ref{eq:poiseulle} for comparison to the macroscopic theory.
As expected for the conductivity, we observe approximately an order of magnitude difference between the peaks in the hydraulic mobilities in the BN and GR nanochannels. 
The mobility profile is nearly flat for GR and exhibits a slight curvature for BN, indicative of the differences in slip. 
In the boundary region, the mobility profile qualitatively mimics the fluid density profile with greater (lesser) flux coinciding with density peaks (troughs). 

We find that the molecular interfacial structure also affects the cross-terms in the mobility matrix in Eq.~\ref{eq:lr}.
The streaming mobility $\mathcal{M}_{jQ}$, which quantifies the 
electrical current profile produced by applying a pressure differential,
is shown in Fig~\ref{fig:mobilities}(b) for both systems. 
We observe the emergence of three layers of electrical current of alternating sign near the fluid wall boundary, and no net current in the bulk of the channel. 
Because the applied pressure produces particle flux in all nanochannel regions, the alternating current is caused by 
ion density localization at the interface, with positive (negative) current where potassium (chloride) ions are enriched. 
These interfacial effects decay away from the wall more slowly than those observed with the hydraulic mobility, 
with net charge flux penetrating into the hydrodynamic region defined by the hydraulic mobility. 
By integrating the mobility across the channel, we find that the streaming conductivity $\mathcal{L}_{JQ}$ is statistically indistinguishable from zero for both materials, indicating no net ionic transport.
Though not shown, our calculations verify the lack of symmetry between cross-term mobilities, with $\mathcal{M}_{qJ}$ being zero at all points in the channel, within statistical accuracy, consistent withq $\mathcal{M}_{qJ}\neq\mathcal{M}_{jQ}$ while maintaining $\mathcal{L}_{QJ}=\mathcal{L}_{JQ}$. 

The pressure driven excess water mobility $\mathcal{M}_{dQ}$, is shown in Fig.~\ref{fig:mobilities}(c) as computed using Eq.~\ref{eq:mobility} for both materials.
This quantity is directly related to the desalination capabilities of a nanofluidic channel, and its magnitude determined by the channel's selectivity and permeability.
This transport is summarized by the integrated mobility, $\mathcal{L}_{dQ}$, with $\mathcal{L}_{dQ}>0$ corresponding to the selective flux of water through the channel. We find a positive integrated value $\mathcal{L}_{dQ}>0$ for both materials, demonstrating a preferential water selectivity and corresponding salt rejection capability. 

The spatial dependence of the cross-term mobility profiles can be understood via a combination of microscopic and macroscopic perspectives. 
The streaming mobility may be evaluated microscopically as a product of the local density profiles and the hydraulic mobility. For the streaming mobility this is, $\mathcal{M}_{jQ}(z)=[\rho_{\mathrm{K}^+}(z)-\rho_{\mathrm{Cl}^+}(z)]/\rho_{\mathrm{tot}}(z) \mathcal{M}_{qQ}(z)N/V  $, where $\rho_\mathrm{tot}(z)=\rho_\mathrm{w}(z)+\rho_{\mathrm{K}^+}(z)+\rho_{\mathrm{Cl}^-}(z)$. Though a common decomposition in macroscopic hydrodynamics, this is a nontrivial statement when considering the microscopic mobilities.
The red dashed line in Fig.~\ref{fig:mobilities}(b) shows this estimate agrees well with estimate using Eq.~\ref{eq:mobility}.
The same functional decomposition holds for the excess water flux, which can be obtained from the product of the hydraulic mobility and the excess water density 
$\mathcal{M}_{dQ}(z)=\left (\rho_\mathrm{w}(z)/\rho_\mathrm{tot}(z)-\bar{\rho}_\mathrm{w}/\bar{\rho}_\mathrm{tot} \right ) \mathcal{M}_{qQ}(z)N/V $. 
This decomposition is shown in the red dashed line in Fig.~\ref{fig:mobilities}(c). Both of these decompositions follow directly from the Langevin equations of motion. 
While the excess water mobilities for both materials are qualitatively similar because of
qualitatively similar equilibrium density distributions and hydraulic mobility profiles, 
the quantitative difference arises due to the differences in magnitude of the hydraulic conductivity. 
The first contact layer is nearly salt free, so while interfacial friction slows pressure driven transport, 
the high water purity gives a large peak in excess water mobility. 
There is a second excess water mobility peak near the second water density peak. The enrichment and depletion of chloride and potassium, respectively, brings the overall salt density close to its bulk value and leaves an excess concentration of water where the hydraulic mobility also peaks.

\begin{figure}
\includegraphics[width=8.5cm]{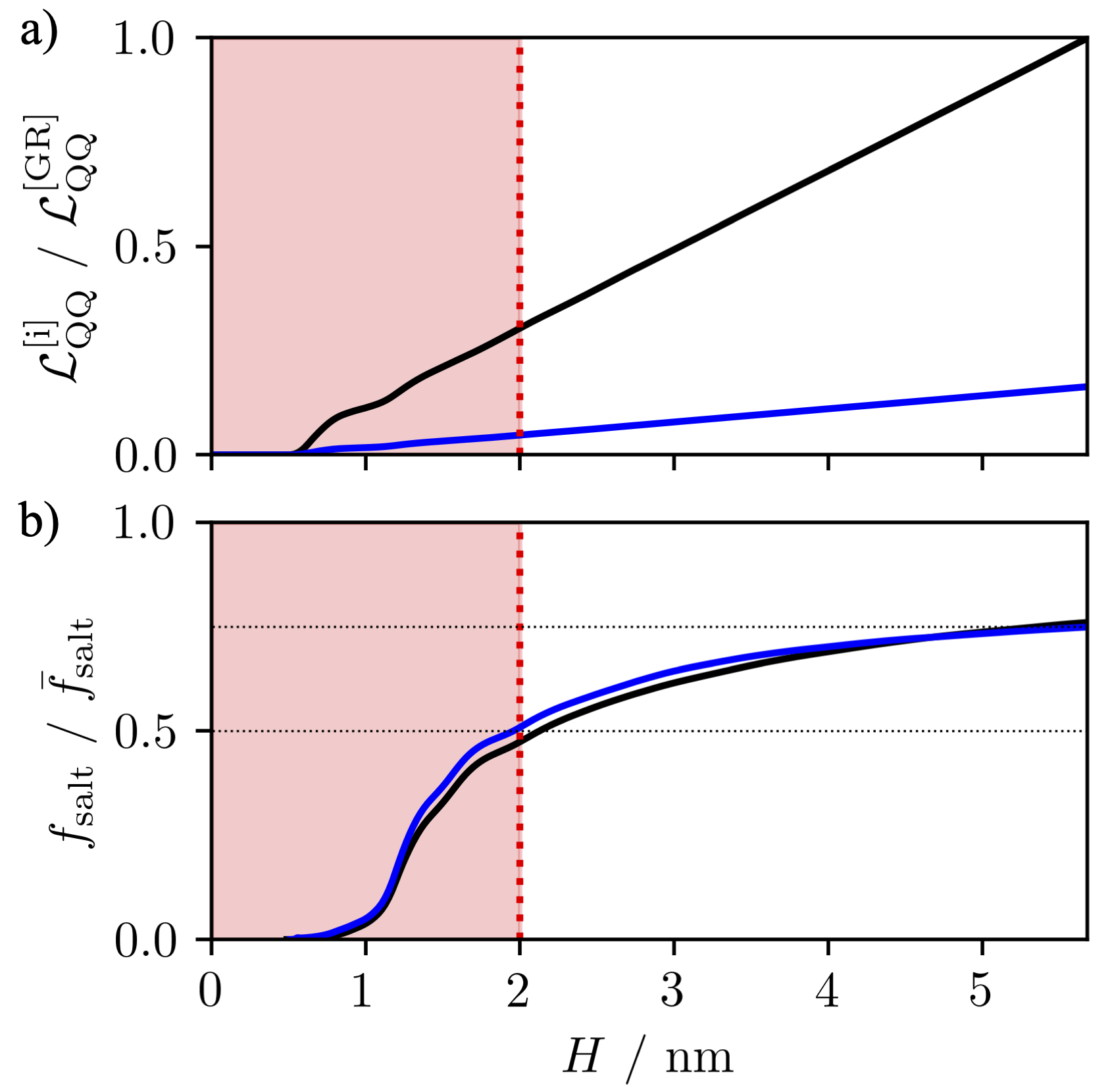}
\caption{Estimates of (a) hydraulic conductivity and 
(b) water selectivity in 
simple GR (black) and BN (blue) nanochannels versus 
channel height $H$. Red shaded regions indicate
channel heights where boundary effects from confining walls interact, 
meaning our estimate is most reliable for 
$H \gtrsim 2 \ \mathrm{nm}$.}
\label{fig:int_mobilities}
\end{figure}

The molecular dynamics calculations suggest that the transport properties of the nanochannel can be decomposed as a sum of a molecular interfacial component, and a continuum bulk component. The interfacial component depends sensitively on specific molecular interactions as they manifest in non-uniform density profiles. Beyond the domain of those density correlations, which for these channels extend around 2 nm into the channel, the transport is well described by Poiseuille flow with a large slip length. This decomposition allows us to infer the height dependence of the channel's selectivity and permeability. 
We can calculate the size dependent conductivity
using an integrated mobility 
$\mathcal{L}_{QQ}({H})=2\int_0^{{H}/2} dz \ \mathcal{M}_{qQ}(z)/H$, where we employ the inversion symmetry of the channel to integrate over only half of the channel. 
These conductivities are shown for BN and GR in Fig.~\ref{fig:int_mobilities}(a) normalized against $\mathcal{L}_{QQ}^\mathrm{[GR]}$. The red regions in Fig. \ref{fig:int_mobilities} indicate system sizes which would lead to overlapping interfacial regions, for which our decomposition is not anticipated to be valid. 
Because the hydraulic mobility profile is nearly flat in the hydrodynamic region, which is expected when $l_s \gg H_\mathrm{hyd}/6$, 
the overall permeability increases linearly with channel height, which is not as fast as anticipated from traditional hydrodynamics with a no-slip boundary condition. 

A similar approach can be used to compute the dependency of the water selectivity on the height of the channel. 
To compute the selectivity, we first can determine a pressure driven salt mobility
$\mathcal{M}_{\mathrm{s}Q}(z)=[\rho_{\mathrm{K}^+}(z)+\rho_{\mathrm{Cl}^-}(z)]/\rho_{\mathrm{tot}}(z)\mathcal{M}_{qQ}(z)N/V$.
The ratio of salt to total particle flux as a function of channel height is obtained as
\ba
f_\mathrm{salt}({H})=\frac{\int_0^{{H}/2} dz \ \mathcal{M}_{\mathrm{s}Q}(z)} {\frac{N}{V}\int_0^{{H}/2} dz \ \mathcal{M}_{qQ}(z)}
\ea
which is shown in Fig.~\ref{fig:int_mobilities}(b) normalized against the overall number fraction of ions in the bulk, 
$\bar{f}_\mathrm{salt}=(\bar{\rho}_\mathrm{K^+}+\bar{\rho}_\mathrm{Cl^-})/\bar{\rho}_{\mathrm{tot}}$.
This  provides a direct measurement of the size dependence of the nanochannel selectivity. Consistent with the inference from the excess water mobility, the salt flux is supressed relative to its expected value from the bulk concentration of ions and the total channel conductivity. 
We find that BN and GR nanochannels have effectively identical selectivities, 
primarily because of their similar equilibrium fluid density distributions and qualitatively similar hydraulic mobility profiles. 
For the nanochannel size and ion concentrations considered here, the flux of salt ions is reduced by approximately 25\%,
while shrinking the nanochannel until interfacial regions overlap at around $2 \ \mathrm{nm}$ could provide a reduction of around 50\%. Due to the intrinsic interfacial absorption of ions to the interface and their resultant suppressed mobility, as the nanochannel size is decreased its selectivity is enhanced. An optimal desalination device must separate ions from water with both high selectivity as well as high permeability, and these phenomenological channel scaling observations suggests that for both BN and GR this optimum is between 2 and 5 nm.

This mechanism of selective transport, and the ability of the channel to separate salt from water, is a result of an interplay between local molecular interactions that drive ions to the fluid-solid boundary in the absence of a net surface charge of the substrate. 
These molecular interfacial features established a nonuniform fluid composition across the channel that, when combined with a spatially resolved evaluation of the hydraulic mobilities, provide a complete description of the transport within the nanochannel. 
The promise of this mechanism for desalination technology is strikingly enhanced when this water selectivity is coupled with the anomalously high permeability of GR nanochannels.  
This framework is general and can be used to understand and engineer other functionality in nanofluidic systems. Employing recent generalizations of response theory,\cite{gao2019nonlinear,lesnicki2020field, lesnicki2021molecular} our  approach could be extended outside the regime of linear response to provide insight into performance at high driving strengths and between multiple driving forces.

{\textbf{\emph{Acknowledgments --}}}
This study is based on the work supported by the U.S. Department of Energy, Office of Science, Office of Advanced Scientific Computing Research, Scientific Discovery through Advanced Computing (SciDAC) program, under Award No. DE-AC02-05CH11231. A. R. P was also supported by the Heising-Simons Fellowship from the Kavli Energy Nanoscience Institute at UC Berkeley and D. T. L acknowledges support from the Alfred P. Sloan Foundation.

{\textbf{\emph{Data availability --}}}  
The source code for the calculations done and all data presented in this work 
are openly available on Zenodo at https://doi.org/10.5281/zenodo.7522996~\sct{zenodo_data}

\bibliography{main}
\end{document}